\documentclass[conference]{IEEEtran}
\IEEEoverridecommandlockouts
\usepackage{cite}
\usepackage{amsmath,amssymb,amsfonts}
\usepackage{algorithmic}
\usepackage{graphicx}
\usepackage{textcomp}
\usepackage{xcolor}

\usepackage{siunitx}
\usepackage{tabularx}

\usepackage{tikz}
\usepackage{hyperref}
\usepackage{lipsum}

\newcommand\copyrighttext{%
  \textcopyright 2019 IEEE. 
  DOI: \href{https://doi.org/10.1109/SGSMA.2019.8784556}{10.1109/SGSMA.2019.8784556}}
\newcommand\copyrightnotice{%
\begin{tikzpicture}[remember picture,overlay]
\node[anchor=south,yshift=10pt] at (current page.south)
{\copyrighttext};
\end{tikzpicture}%
}

\newenvironment{conditions}
  {\par\vspace{\abovedisplayskip}\noindent\begin{tabular}{>{$}l<{$} @{${}={}$} l}}
  {\end{tabular}\par\vspace{\belowdisplayskip}}

\def\BibTeX{{\rm B\kern-.05em{\sc i\kern-.025em b}\kern-.08em
    T\kern-.1667em\lower.7ex\hbox{E}\kern-.125emX}}
\begin{document}

\title{Mode Shape Estimation using Complex Principal Component Analysis and k-Means Clustering\\
}

\author{\IEEEauthorblockN{Hallvar Haugdal and Kjetil Uhlen, \textit{Member, IEEE}}
\IEEEauthorblockA{{Department of Electric Power Engineering,} {NTNU} \\
Trondheim, Norway \\
}
}

\maketitle
\copyrightnotice

\begin{abstract}

%
We propose an empirical method for identifying low damped modes and corresponding mode shapes using frequency measurements from a Wide Area Monitoring System. The method consists of two main steps: Firstly, Complex Principal Component Analysis is used in combination with the Hilbert Transform and Empirical Mode Decomposition to provide estimates of modes and mode shapes. The estimates are stored as multidimensional points. Secondly, the points are grouped using a clustering algorithm, and new averaged estimates of modes and mode shapes are computed as the centroids of the clusters.
%
%
%
%
Applying the method on data resulting from a non-linear power system simulator yields estimates of dominant modes and corresponding mode shapes that are similar to those resulting from modal analysis of the linearized system model. Encouraged by the results, the method is further tested with real PMU data at transmission grid level. Initial results indicate that the performance of the proposed method is promising.

\end{abstract}


\section{Introduction}
Power system operators are facing the increasingly evident challenge of operating the grid securely during complex, uncertain and rapidly changing generation and demand. This challenge can and must be met by providing the operators with better tools for monitoring and control of the grid. In this paper we focus on the monitoring part and how the operators can improve their situational awareness and gain better knowledge about stability properties and dynamic phenomena occurring in the grid.



Worldwide, Phasor Meaurement Units (PMUs) are being installed within transmission grids at increasingly faster rates. It is expected that the information from PMUs will replace the current remote terminal units (RTUs) as source of voltage and current measurements to SCADA/EMS systems. This will change the way operators can monitor the system state. First of all, Wide Area Monitoring Systems opens new possibilities for getting fast and precise information about system dynamics. While WAMS have been in use many places for offline studies and fault analysis, the development of applications for online use has not yet matured.
One reason for this may be that it takes time for operators to get used to and take advantage of the new technology. Research is therefore needed in order to develop applications that are readily accepted and understood by the operators. This paper aims to contribute to that development.

The method proposed in this paper takes advantage of the high number of frequency (or potentially voltage angle) measurements available in the transmission grid in order to extract precise information about low damped oscillatory modes and their observability.

Multiple empirical methods for estimating information about modes have been proposed in the litterature. In\cite{Duong2018}, Prony analysis is used to extract frequencies and damping ratios from measured oscillations. This method, however, does not provide information about mode shapes. In \cite{Kumar2017}, a method for estimating modes and mode shapes using various mode decomposition techniques and Power Spectral Density analysis is proposed. In \cite{Seppanen2017}, a Bayesian approach is proposed for monitoring of electromechanical modes.

The proposed method presented in this paper is based on using Principal Component Analysis (PCA) and k-Means clustering for estimating modes and corresponding mode shapes from PMU frequency measurements. The method consists of two main parts: In Part I, the oscillatory behavior measured in a number of measurements is decomposed into a few main oscillatory components using a variant of Complex Principal Component Analysis (CPCA). The decomposition allows the oscillatory behavior to be reduced to a set of parameters, which can be stored in the form of a multidimensional point, referred to as an observation. Finally, in Part II of the method, the observations from Part I are clustered using the k-Means algorithm, to determine which of the observations to include when computing averaged modes and mode shapes.

CPCA is well described in \cite{Horel1984}, where it is aimed at extracting dynamic patterns in geophysical data sets. Variants of CPCA has been applied to power system analysis in\cite{Messina2007,P2009,Messina2011}. The implementation used in this paper is slightly different, but functions in much the same way. An important contribution from this paper is the addition of the additional layer, referred to as Part II of the method, where the information provided by CPCA is processed further into more easily understandable information, i.e. by filtering and clustering observations and computing averaged quantities.



We present and describe the necessary background theory in section \ref{sec:Theory}, and describe the proposed method in section \ref{sec:ProposedMethod}. Application examples on simulated as well as real PMU data are presented in section \ref{sec:Application}. Finally, discussion and conclusions are given in sections \ref{sec:Discussion} and \ref{sec:Conclusion}.


\section{Background theory}
This section provides a short introduction to modal analysis, and presents theory on the most important tools used by the proposed method.
\label{sec:Theory}
\subsection{Modal Analysis}

Modal analysis of power systems provides useful knowledge of the system dynamics
, and can be used for multiple purposes, including assessment of small signal stability, and design and tuning of power oscillation dampers \cite{Johansson2009}.

Fundamental to modal analysis of power systems is the linearization of the dynamic equations of the system. The state space form, assumed familiar to the reader, is given as follows\cite{Kundur1994}:
\begin{equation}
\Delta \mathbf{\dot{x}} = \mathbf{A} \Delta \mathbf{x} + \mathbf{B} \Delta \mathbf{u}
\end{equation}

Through eigendecomposition of the system matrix $\mathbf{A}$, the dynamic modes of the system are determined from the eigenvalues, and the observability of the modes (mode shapes) are contained within the elements of the right eigenvectors. It can be shown that the response of the system can be written

\begin{equation}
\label{eq:Observability}
\begin{split}
    \Delta \mathbf{x}(t) &= \sum_{j=1}^n \mathbf{\Phi}_j c_j e^{\lambda_j t}\\
\end{split}
\end{equation}
where
\begin{conditions}
\mathbf{\Phi}_j	&  right eigenvector $j$ of system matrix\\
c_j 	&  magnitude of excitation of the $j$th mode\\   
\lambda_j & eigenvalue $j$ of system matrix\\
n & order of the system
\end{conditions}

The element $\Phi_{ij}$ of $\mathbf{\Phi}_j$ describes the amplitude and phase with which mode $j$ appears in state $\Delta x_i$.


When performing modal analysis analytically, difficulties arise when modelling large systems due to insufficient knowledge of the system topology, component characteristics and controller tuning. Therefore, methods are developed for online identification of electromechanical modes that are based on measurements only. However, whether the methods are model based utilizing a state estimator or purely based on measurements, the concept of modes is inherently related to linear theory. Power systems, however, are nonlinear by nature and what appears as critical modes may change rapidly during a disturbance or any changes in operation.


\subsection{Principal Component Analysis}
Principal Component Analysis (PCA, sometimes referred to as Empirical Orthogonal Functions) is a powerful dimensionality reduction technique, which allows the variance in a set of time series to be decomposed into a few orthogonal Principal Components (PCs) that explain the main share of the variance.
A brief introduction of the method is given below, based on the thorough description of the method given in \cite{Jackson1991}.

Assuming we have $M$ measurement series, each containing $N$ samples, assembled in a $M\times N$ matrix as follows:
\begin{equation}
\mathbf{X}=
\begin{bmatrix}
\mathbf{x}_1\\
\mathbf{x}_2\\
\vdots\\
\mathbf{x}_M
\end{bmatrix}
=
\begin{bmatrix}
x_{1}(t_1) & x_{1}(t_2)& \cdots & x_{1}(t_N)\\
x_{2}(t_1) & x_{2}(t_2) &\cdots & x_{2}(t_N)\\
\vdots & &\ddots & &\\
x_{M}(t_1) & x_{M}(t_2) & & x_{M}(t_N)
\end{bmatrix}
\end{equation}

Here, the notation $x_i(t_k)$ denotes the value of series $i$ at time $t_k$. We assume that each of the series has zero mean (if this is not the case, the mean is subtracted prior to the analysis).

We want to transform the correlated series $\mathbf{x}_1,\mathbf{x}_2\dots\mathbf{x}_M$ into the uncorrelated series $\mathbf{s}_1,\mathbf{s}_2\dots\mathbf{s}_M$ using the the linear transformation

\begin{equation}
\mathbf{S}=
\begin{bmatrix}
\mathbf{s}_1\\
\mathbf{s}_2\\
\vdots\\
\mathbf{s}_M
\end{bmatrix}
=
\mathbf{U}^T\mathbf{X}
\label{PCA}
\end{equation}

To find the matrix $\mathbf{U}$, we start by establishing the correlation matrix of $\mathbf{X}$:


\begin{equation}
\begin{split}
\mathbf{C}= \frac{1}{1-N} \mathbf{X} \mathbf{X}^T
\Rightarrow \left( 1-N \right) \mathbf{C}= \mathbf{X} \mathbf{X}^T
\end{split}
\end{equation}

Similarly, the covariance matrix of $\mathbf{S}$ is given by 


\begin{equation}
\begin{split}
\left( 1-N \right)\mathbf{\Lambda}=& \mathbf{S} \mathbf{S}^T\\
=& \mathbf{U}^T\mathbf{X}\left(\mathbf{U}^T\mathbf{X}\right) ^T\\
=& \mathbf{U}^T\mathbf{X}\mathbf{X}^T\mathbf{U}\\
=& \mathbf{U}^T\mathbf{C}\mathbf{U}
\end{split}
\end{equation}

Since we want the series $\mathbf{S}$ to be uncorrelated, we want the matrix $\mathbf{\Lambda}$ to have non-zero elements only along the diagonal. Thus, we want to choose the matrix $\mathbf{U}$ such that $\mathbf{C}$ is diagonalized. The diagonalization is carried out by performing an eigendecomposition of $\mathbf{C}$, where the eigenvalues and eigenvectors are given by

\begin{equation}
\mathbf{C}\mathbf{u}_j=\lambda_j \mathbf{u}_j, j=1\dots M
\end{equation}

The relative magnitudes of the eigenvalues describe how much variance the corresponding PC explains. The eigenvalues are sorted based on magnitude in descending order and collected in the eigenvalue matrix $\mathbf{\Lambda}$, and the eigenvectors are collected as column vectors in the transformation matrix $\mathbf{U}$ in the same order. Finally, this gives us the $M$ uncorrelated series $\mathbf{S}$, which can be referred to as the Scores of the PCs. If the input series $\mathbf{X}$ are highly correlated, then we can represent a high share of the variance using only a low number $M_{PC}<<M$ of the first PCs.

We can also invert the transformation in \eqref{PCA}. Since it can be shown that $\mathbf{U}$ is an orthonormal matrix \cite{Jackson1991}, the inverse of the matrix $\mathbf{U}$ is simply the transpose of the matrix:
\begin{equation}
\begin{split}
&\mathbf{X}=(\mathbf{U}^T)^{-1} \mathbf{S}=\mathbf{U}^{} \mathbf{S}\\
&\Rightarrow \mathbf{x}_i = \sum_{j=1}^{M} u_{ij}\mathbf{s}_j \approx \sum_{j=1}^{M_{PC}}u_{ij}\mathbf{s}_j
\end{split}
\label{PCcontribution}
\end{equation}

From the above relation we can state that PC $j$ appears in measurement $i$ with a magnitude given by the coefficient $u_{ij}$.








\subsection{Complex Principal Component Analysis}
One limitation with conventional PCA is that a travelling wave, occuring in multiple measurements with different phase, can not be captured by only one PC. However, an extended version of PCA, often referred to as Complex Principal Component Analysis (CPCA), is better suited for this purpose. The method, described in detail in \cite{Horel1984}, uses the same equations as conventional PCA presented above, but some additional steps are required prior to the analysis.

CPCA is, in short, the same equations as presented above applied to complex time series, where the real part of the series is the input series, and the complex part is generated by applying the Hilbert Transform. To get sensible results when using the Hilbert Transform, it is important to make sure that the signal to be transformed contains only one frequency, and that any non-oscillatory trends are subtracted. This can be achieved using Empirical Mode Decomposition (EMD), or by applying PCA in two layers, discussed further in section \ref{sec:PCA-CPCA}.

The Hilbert Transform is also strongly influenced by end effects \cite{Horel1984}. When taking the transform of a measured series, this often results in very high values at the beginning and end of the series. To avoid that this influences consecutive steps in the algorithm, roughly \SI{10}{\%} of the series should be tapered in each end\cite{Horel1984} after taking the Hilbert Transform.


After applying the necessary pre-processing to ensure a well-posed Hilbert Transform, the complex series can be written on the form

\begin{equation}
\label{eq:Hilbert}
\mathbf{y}_i=\mathbf{x}_i+jH\left(\mathbf{x}_i\right)
\end{equation}
where $H(\cdot)$ denotes the Hilbert Transform, and $\mathbf{x}_i$ is the real input series (with zero mean).

The complex series are collected in a matrix,
\begin{equation}
\mathbf{Y}=
\begin{bmatrix}
\mathbf{y}_1\\
\mathbf{y}_2\\
\vdots \\
\mathbf{y}_M
\end{bmatrix}
\end{equation}

The complex covariance matrix is given by
\begin{equation}
\left( 1-N \right) \mathbf{\tilde{C}}= \mathbf{Y} \mathbf{Y}^H
\end{equation}
where the superscript $(\cdot)^H$ denotes the conjugate transpose (not to be confused with the Hilbert Transform). Diagonalizing results in real eigenvalues\cite{Horel1984} and complex eigenvectors. Analogously to \eqref{PCcontribution}, we have
\begin{equation}
\begin{split}
&\mathbf{Y}=(\mathbf{V}^{H})^{-1} \mathbf{Z}=\mathbf{V}^{} \mathbf{Z}\\
&\Rightarrow \mathbf{y}_i = \sum_{j=1}^{M} v_{ij}\mathbf{z}_j \approx \sum_{j=1}^{M_{CPC}} v_{ij}\mathbf{z}_j
\end{split}
\label{CPCcontribution}
\end{equation}
where $\mathbf{V}$ is the matrix of complex eigenvectors $\mathbf{v}_i$, with elements $v_{ij}$, and $\mathbf{Z}$ contains the Scores of the CPCs $\mathbf{z}_j$ as column vectors. The complex components of the eigenvectors describe the amplitude and phase of the contribution of each of the CPCs to each of the measurements. Here, the correlation matrix is Hermitian \cite{Horel1984}, the matrix $\mathbf{V}$ is unitary, and thus its inverse is the conjugate transpose.

\subsection{Two-layer combination of PCA and CPCA}
\label{sec:PCA-CPCA}
Applying the Hilbert Transform requires the input signal to have only one frequency, and that any non-oscillatory trends are subtracted. This can be achieved by using EMD to decompose the input signal into Intrinsic Mode Functions (IMFs) and a Residual\cite{Messina2011}. The Hilbert Transform can then be applied to the IMFs, before CPCA is applied on the resulting complex series.

An alternative to using EMD for decomposition into IMFs, is to apply a to a two-layer structure of PCA: In the first layer conventional PCA is applied on the input series, resulting in a number of PCs. In the second layer, the PCs from the first layer are detrended by subtracting the residual from EMD, the Hilbert Transform is applied, before finally CPCA is applied. Thus, EMD is used only for computing the non-oscillatory residual, not for decomposition into IMFs. Detrending by EMD is described in \cite{Flandrin2015}.

In equations we can write, for the first layer (where PCA is applied),

\begin{equation}
    \mathbf{S}=\mathbf{U}^T\mathbf{X}
\end{equation}

The scores from the first layer are detrended (i.e. the residual computed using EMD is subtracted),

\begin{equation}
    \mathbf{S}'=\mathbf{S}-\mathbf{R}
\end{equation}
the Hilbert Transform is applied,
\begin{equation}
    \mathbf{Y}=\mathbf{S}'+jH\left(\mathbf{S}'\right)
\end{equation}
(Here, $H(\mathbf{S}')$ denotes series-wise application of the Hilbert Transform, similarly as in \eqref{eq:Hilbert}. Also, although not shown with equations here, in a practical implementation roughly \SI{10}{\%} should be tapered in each end at this stage, to limit end effects.)

This gives the complex series which constitutes the input of the second layer (where CPCA is applied),
\begin{equation}
        \mathbf{Z}=\mathbf{V}^H\mathbf{Y}
        =\mathbf{V}^H\left(\mathbf{S}'+jH(\mathbf{S}')\right)
\end{equation}

Combining the above equations, we can write
\begin{equation}
    \begin{split}
        \mathbf{X}&=\mathbf{U}\mathbf{S}\\
        &=\mathbf{U}\left(\mathbf{S}'+\mathbf{R}\right)\\
        &=\mathbf{U}\left(\mathrm{Re}\left(\mathbf{Y} \right) + \mathbf{R}\right)\\
        &=\mathbf{U}\left(\mathrm{Re}\left(\mathbf{V Z} \right) + \mathbf{R}\right)\\
    \end{split}
\end{equation}

Neglecting the residual and defining the matrix $\mathbf{W}=\mathbf{UV}$, we get
\begin{equation}
    \mathbf{X}=\mathrm{Re}\left(\mathbf{UV Z} \right) =\mathrm{Re}\left(\mathbf{W Z} \right)
\end{equation}

Finally, we can write

\begin{equation}
\mathbf{x}_i = \sum_{j=1}^{M} w_{ij}\mathbf{z}_j \approx \sum_{j=1}^{M_{CPC}}w_{ij}\mathbf{z}_j
\label{eq:CoefficientsFinal}
\end{equation}

This gives the contribution of the CPC $\mathbf{z}_j$, to the measurement $\mathbf{x}_i$ as the coefficient $w_{ij}$, i.e. element $(i,j)$ of the matrix $\mathbf{W}$.

Note that if no components are discarded in either of the layers, then the matrices $\mathbf{U}$, $\mathbf{V}$ and $\mathbf{W}$ will be $M\times M$-matrices. If components are discarded, such that only $M_{PC}$ components are kept from the first layer and $M_{CPC}$ from the second layer, $\mathbf{U}$, $\mathbf{V}$ and $\mathbf{W}$ will be of dimension $M\times M_{PC}$, $M_{PC}\times M_{CPC}$ and $M\times M_{CPC}$, respectively.

Comparing modal analysis with CPCA as described, we see that the coefficients $w_{ij}$ in \eqref{eq:CoefficientsFinal} become very similar to the observability phasors $\phi_{ij}$ in \eqref{eq:Observability}. Thus, if the decomposition successfully produces CPCs with frequency and damping similar to that of the modes of the system, then the coefficients $w_{ij}$ can be considered estimates of observability mode shapes.

\subsection{k-Means Clustering}
\label{sec:Clustering}
Clustering, in general, refers to the task of dividing a set of observations into groups, such that observations belonging to the same group are relatively similar, while the observations belonging to different groups are relatively different \cite{James2013}. In the proposed method, clustering is used to determine which mode estimates, or observations, are similar enough to be included when computing averages of different modes and mode shapes. The k-Means clustering algorithm is widely used due to its computational simplicity. However, the method is prone to noise \cite{Theodoridis2009}, and requires the number of clusters to be determined before the clustering.


To determine the number of clusters, a simple approach is to carry out the clustering multiple times with a range of different numbers of clusters, and then determining the most optimal solution by assessing some measure of the "goodness" of each clustering. The \textit{Silhouette index} can be adopted for this purpose. 
For a thorough description of the k-Means algorithm and formal definitions of the silhouette index, the interested reader is referred to \cite{Theodoridis2009}.


\subsection{Exponential fitting by Linear Regression}
\label{sec:Regression}
Linear regression can be used for estimating the decay rate of an exponentially decaying series. We would like to describe the data $Y_i$ measured at $X_i$ using the exponential function $Y_i=\alpha e^{\beta X_i}+\varepsilon$, where $\varepsilon$ is the error in the estimate. The parameters $\alpha$ and $\beta$ are estimated by taking the logarithm of the values $Y_i$ in the series, and then using Simple Linear Regression to find the linear relationsship between $X_i$ and $\ln(Y_i)$. Expressions for $\alpha$ and $\beta$ are given in \cite{Glaister2007}.

Once we know the parameters, we can compute the average squared error by $
R=\frac{1}{N} \sum_{i=1}^N \left(\alpha e^{\beta X_i}-Y_i\right)^2
$.

\section{Proposed method}
\label{sec:ProposedMethod}
 The two main parts of the proposed method are described in detail below; in short, the purpose of the first part of the method is to provide numerous estimates of modes using dimensionality reduction techniques, while in the second part averaged estimates of modes and mode shapes are computed using clustering. We assume $M$ frequency measurement streams from which we want to extract information about modes and mode shapes.

\subsection*{Part I - Mode estimation}
\label{PartA}

The first part of the method acts on a sliding time window containing $N$ samples (for each of the $M$ series). The length of the sliding window should be sufficiently long to capture low frequency oscillations
, i.e. in the range $5-\SI{10}{\s}$.

For each time window, we would like to decompose the oscillatory behaviour into a number of components, and find the amplitude and phase with which this behaviour happens in each measurement.
Each time window is analyzed according to the following sequence:

\begin{enumerate}
     \item The two-layer combination of PCA and CPCA, descibed in section \ref{sec:PCA-CPCA}, is applied to the $M\times N$ matrix of input series:
     \begin{enumerate}
        \item Each of the time series are normalized such as to have zero mean. 
        \item PCA is applied,
        \item The scores are detrended,
        \item The Hilbert Transform is applied, and {10}{\%} of the ends of the resulting complex series are tapered
        \item CPCA is applied.
     \end{enumerate}
     \item The frequency and decay rate of the resulting CPCs are estimated. The frequency is computed using the built-in \texttt{MATLAB}-function \texttt{meanfreq()}. The decay rate is computed using exponential regression on the absolute value of the CPCs ($\mathrm{Abs}(\mathbf{z}_i)=\sqrt{\mathrm{Re}(\mathbf{z}_i)^2+\mathrm{Im}(\mathbf{z}_i)^2}$), described in \ref{sec:Regression}.
     \item To limit the amount of uninteresting observations and noise in the clustering, only the estimates with a frequency between \SI{0.1}{Hz} and \SI{2}{Hz} (in the range of electromechanical oscillations) and a regression error below $4\cdot 10^{-3}$, computed as described in section \ref{sec:Regression}, are used in the further analysis. The CPCs passing this criterion, along with the corresponding coefficients describing the amplitude and phase of their presence in the measurements, are expected to be relatively good estimates of modes and mode shapes.
     \item Each of the accepted CPCs are rotated such that the largest coefficient, corresponding to the longest observability phasor in the mode shape, is at \SI{0}{\degree}.
    \item Finally, each mode estimate and the corresponding mode shape is stored as a $2M+2$-dimensional point on the form
    \begin{equation}
\begin{split}
\mathbf{p}=[f,\sigma,\mathrm{Re} (G_1),\mathrm{Im}(G_1),\mathrm{Re} (G_2),\mathrm{Im}(G_2)...\\
...\mathrm{Re} (G_M),\mathrm{Im}(G_M) ]
\label{PointFormat}
\end{split}
\end{equation}
where
\begin{conditions}
f & frequency\\
\sigma & decay rate\\
\mathrm{Re}(\cdot) & real part\\
\mathrm{Im}(\cdot) & imaginary part\\
G_i & observability phasor for measurement $i$
\end{conditions}



\end{enumerate}

Once we have the mode estimates, it does not matter from which CPCs they were derived. Assuming that CPC $j$ fulfilled the criterions in step 3, then the observability phasor of measurement $i$ would be given by the coefficient $w_{ij}$ in \eqref{eq:CoefficientsFinal}, i.e., we have $G_i=w_{ij}$. In the following, the stored points on the form \eqref{PointFormat} are referred to as "observations", to avoid confusion with the final, averaged mode estimates resulting from Part II, which are referred to as "mode estimates".



Step 3 in the sequence above is included to filter out bad observations, due to the k-Means algorithm being susceptible to noise. If a CPC resulting from the two-layer PCA-decomposition successfully captures a mode, then the absolute value of the CPC should decay (or grow) exponentially. Applying regression as described in section \ref{sec:Regression} gives an estimate of the decay rate of the mode, along with the average error in the estimate. A large error implies that the CPC does not decay or grow exponentially, and therefore is probably not a good estimate of one of the modes of the system. The threshold value of $4\cdot 10^{-3}$ is found empirically, resulting in reasonable balance between acceptance and rejection of observations.



\subsection*{Part II - Computing mode averages}
\label{Part2}

The PCA-decomposition in Part I is able to produce snapshots of modes and mode shapes, but with some uncertainty. To digest the numerous observations and provide more reliable estimates, we would like to compute averaged modes- and mode shapes. 

To determine which of the observations to include when averaging each mode, we apply a clustering algorithm: Observations that are similar are be grouped together, and the different groups/clusters are assumed to be associated with different modes. The averaged mode shapes are then computed as the centroids of the clusters.

Assuming we have $Q$ point estimates resulting from Part I, these can be organized in a $Q\times (2M+2)$-matrix:

\begin{equation}
\begin{split}
\mathbf{P}=
\begin{bmatrix}
\mathbf{p}_1\\
\mathbf{p}_2\\
\vdots\\
\mathbf{p}_Q
\end{bmatrix}
=
\begin{bmatrix}
f& \sigma & \mathrm{Re} (G_1)&\dots&\mathrm{Im}(G_M)\\
f& \sigma & \mathrm{Re} (G_1)&\dots&\mathrm{Im}(G_M)\\
\vdots\\
f& \sigma & \mathrm{Re} (G_1)&\dots&\mathrm{Im}(G_M)
\end{bmatrix}
\end{split}
\end{equation}

Clustering of the points is carried out using the k-Means algorithm as described in \ref{sec:Clustering}: The data is clustered multiple times with different numbers of clusters (for instance 1 to 10 clusters), and the final clustering is chosen to be the one with the highest silhouette index.

Finally, the averaged modes are derived from the centroids of the clusters.




\section{Application}
\label{sec:Application}

The method is tested on simulated data from The Kundur Two Area System, performed in DigSILENT PowerFactory, and on PMU data recorded during an oscillatory event in the Nordic Power System. In the case with simulated data, the analysed time series has very high modal content and no ambient noise, and should be regarded as a demonstrational example. The case with PMU data contains ambient noise, and demonstrates the applicability of the method to measurements from a real system.

\subsection{Simulated data from the Kundur Two Area System}

\begin{figure}
  \centering \includegraphics[width=0.45\textwidth]{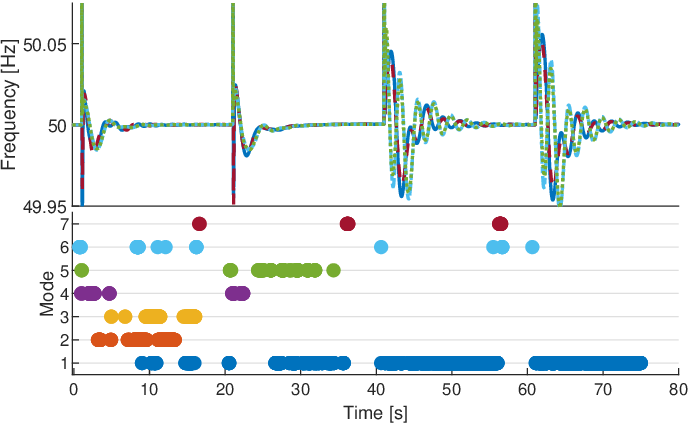}
  \caption{Short circuits with clearing time \SI{20}{ms} at four different locations are simulated in the Kundur Two Area System using DigSILENT PowerFactory, resulting in the time series shown in the upper plot. The lower plot shows where in time the observations contributing to the different mode estimates are detected. The same colors are used in Figs. \ref{fig:K2A-FHist} to \ref{fig:K2A-Modes} to indicate the cluster division.}
  \label{fig:K2A-TimeSeries}
\end{figure}


This system is widely used for studying local and interarea modes. 
The system parameters are described in \cite{Kundur1994}, but the model is modified according to the following: Generators 1 \& 2 emulate hydro power plants, with governor model \texttt{HYGOV}, inertia constants $H=\SI{10}{s}$ and synchronous reactances $X_d=1.2$ \& $X_q=0.9$. Generators 3 \& 4 emulate thermal power plants, with governor model \texttt{TGOV1}. All generators are equipped with the excitation system model \texttt{SEXS}. Finally, the load flow is modified, in order to achieve lower damping of the interarea mode, such that Generators 1, 2, 3 and 4 produce \SI{600}{MW}, \SI{500}{MW}, \SI{801}{MW} and \SI{900}{MW}, respectively. Other parameters are as described in \cite{Kundur1994}.

Short circuits with a clearing time of \SI{20}{ms} are applied subsequently near each of the four generators, with \SI{20}{s} between each short circuit. The resulting frequencies (measured at the generator buses), are shown in Fig. \ref{fig:K2A-TimeSeries}.

\begin{figure}
	\centering \includegraphics[width=0.45\textwidth]{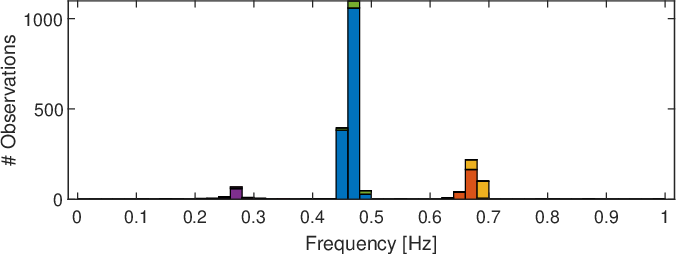}
	\centering \includegraphics[width=0.45\textwidth]{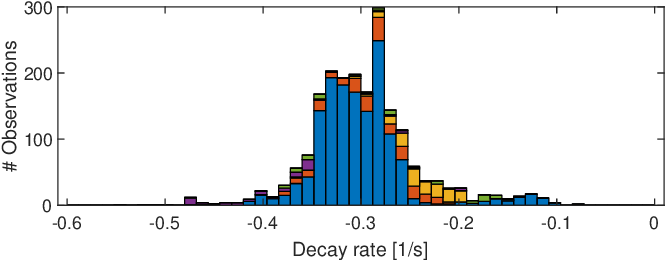}
	\caption{The two histograms indicate the number of observations on the y-axis, and the frequency (upper plot) and decay rate (lower plot) on the x-axis, where the colors indicate the share of each column belonging to each cluster. The same colors as in Fig. \ref{fig:K2A-TimeSeries} to indicate the cluster division.}
  	 \label{fig:K2A-FHist}
\end{figure}

\begin{figure}
	\centering \includegraphics[width=0.41\textwidth]{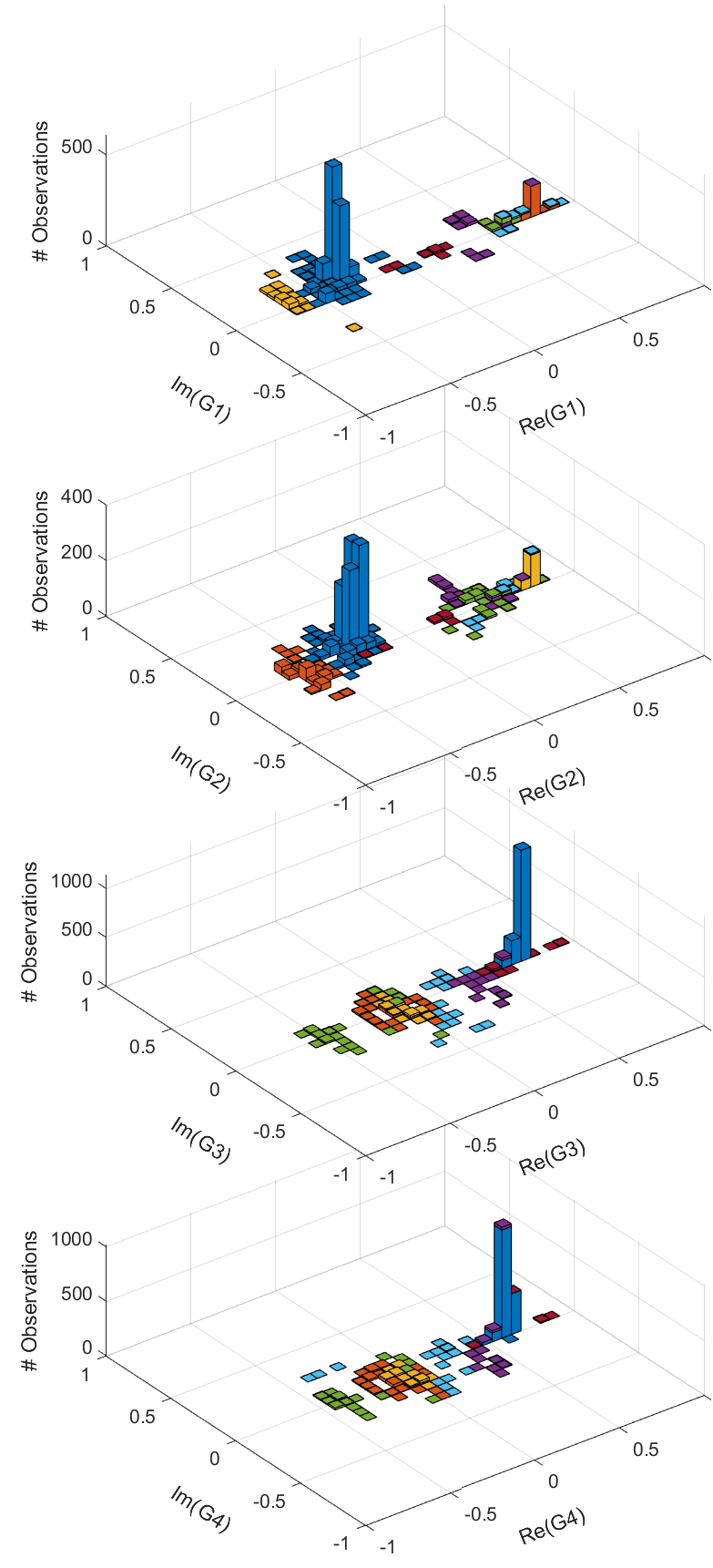}
	\caption{The plots above show 2D histograms indicating the number of occurences (z-axis) of real and imaginary parts (x and y-axis) of the observability phasors of the four generators (first plot = Generator 1, second plot = Generator 2, and so on.). The colors indicate the share of each column belonging to each cluster. The same colors as in Figs. \ref{fig:K2A-TimeSeries} and \ref{fig:K2A-FHist} are used to indicate the cluster division.}
	 \label{fig:K2A-CHist}
\end{figure}

Figs. \ref{fig:K2A-FHist} and \ref{fig:K2A-CHist} show histograms resulting from Part I of the method, indicating the number of observations with a given frequency, damping and observability phasor for each generator. Considering the histogram for frequency in Fig. \ref{fig:K2A-FHist}, we see that higher densities of observations form around frequencies $0.3, 0.5$ and \SI{0.7}{Hz}. 

Part II of the method performs the clustering of the densely populated areas, and results in 7 clusters. The lower plot in Fig. \ref{fig:K2A-TimeSeries} shows the time instants where observations assigned to the different clusters are obtained.

\begin{figure}
  \centering \includegraphics[width=0.45\textwidth]{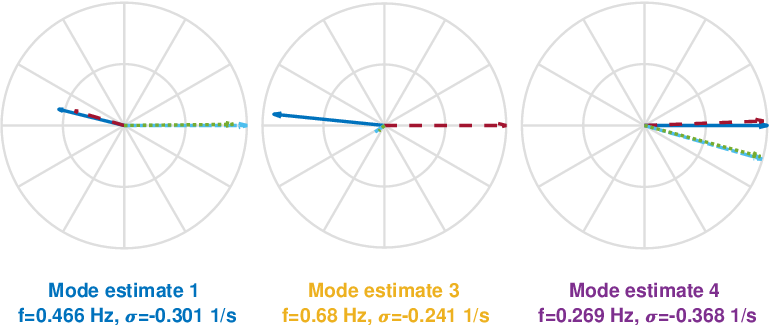}\\
    \vspace{0.4cm}
      \centering \includegraphics[width=0.45\textwidth]{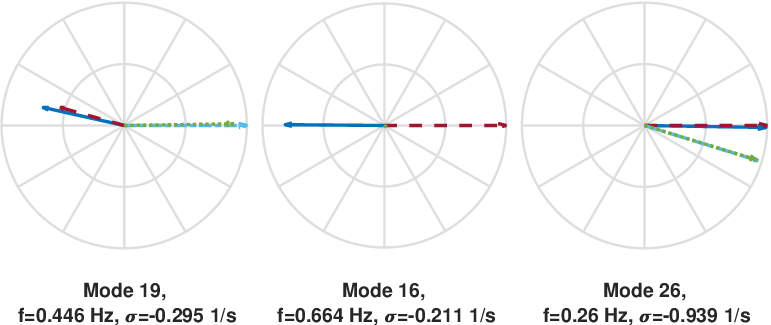}
  \caption{The three upper mode shapes result from applying the method to simulated time series from the Kundur Two Area System, while the lower three mode shapes are computed using modal analysis on a linearized model of the same system. In the colored captions of the upper three mode shapes, the same colors as in Figs. \ref{fig:K2A-TimeSeries} to \ref{fig:K2A-CHist} are used to indicate the cluster division. For the phasors in all the mode shapes, the colors and line types correspond to those of the time series in Fig. \ref{fig:K2A-TimeSeries}.}
    \label{fig:K2A-Modes}
\end{figure}

The coloring of the columns of the histograms indicate the share of the observations being assigned to different clusters, for instance, most of the $\approx\SI{0.5}{Hz}$-observations are assigned to the cluster shown in blue, and most of the $\approx\SI{0.7}{Hz}$-observations are assigned to the cluster shown in red, and so on. The same colors are used in the lower part of Fig. \ref{fig:K2A-TimeSeries} and Figs. \ref{fig:K2A-FHist} to \ref{fig:K2A-Modes} to indicate the clustering division. Considering the observations around \SI{0.5}{Hz} in Fig. \ref{fig:K2A-FHist}, shown in blue, we can study Fig. \ref{fig:K2A-CHist} to find that in these observations, the observability phasors for generators 1 \& 2 (shown in the two upper plots) in average appear to have a real part around -0.4, and an imaginary part of about zero. For generators 3 \& 4, the corresponding real and imaginary parts appear to be situated around 0.7 and zero. This indicates that this is an interarea mode, where generators 1 \& 2 are swinging against generators 3 \& 4. Considering the lower plot in Fig. \ref{fig:K2A-FHist}, the decay rate of the same mode appears to be less accurately estimated due to the high variance compared to the other histograms, but the average appears to be around \SI{0.3}{1/s}.

Once the clustering is performed, we know which observations to include when averaging each mode. The averaged modes from three of the clusters are shown in the upper part Fig. \ref{fig:K2A-Modes}. For comparison, the corresponding modes computed using modal analysis in \texttt{DigSILENT PowerFactory} are shown below, indicating that the method is capable of providing accurate estimates of system modes. The most significant deviation is the decay rate of the third mode shown, where the proposed method produces an estimate of \SI{-0.37}{1/s}, while modal analysis gives the value \SI{-0.939}{1/s}. Note that this mode is nevertheless very well damped, and that it is a mode associated with turbine governors. The method is thus not limited to identification of electromechanical modes.

As mentioned, the method results in 7 clusters, representing 7 averaged modes, whereof three are shown in \ref{fig:K2A-Modes}. By comparing with results from modal analysis, it is found that two of the averaged mode estimates, number 6 and 7, do not resemble any of the modes from modal analysis. 

Further, it is found that mode estimate 2 and 3 are very similar, except that the mode shape is rotated \SI{180}{\degree}, which is probably due to the mode shapes from Part I being rotated such that the longest observability phasor lies at \SI{0}{\degree}. Due to some inaccuracy in the CPCA decomposition, the next longest observability phasor will in some estimates appear the longest, thus flipping the whole mode shape around. The same is the case for estimates 1 and 5, the averaged mode shapes are very similar, but appear at different angular positions.


\subsection{PMU measurements from oscillatory event}


\begin{figure}
  \centering \includegraphics[width=0.25\textwidth]{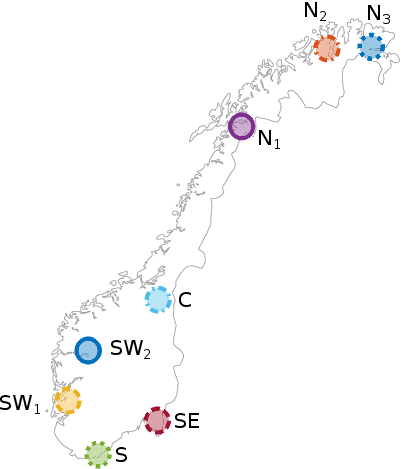}
  \caption{This map shows the approximate locations of the eight PMUs recording the oscillations shown in Figs. \ref{fig:PMU-TimeSeries} and \ref{fig:PMU-TimeSeries-Excerpt}. The appearance of the markers correspond to the colors and line types of the time series in Figs. \ref{fig:PMU-TimeSeries}, \ref{fig:PMU-TimeSeries-Excerpt} and the phasors in Fig. \ref{fig:PMU-Modes}.}
  \label{fig:NorwayMap}
\end{figure}

\begin{figure}
  \centering \includegraphics[width=0.45\textwidth]{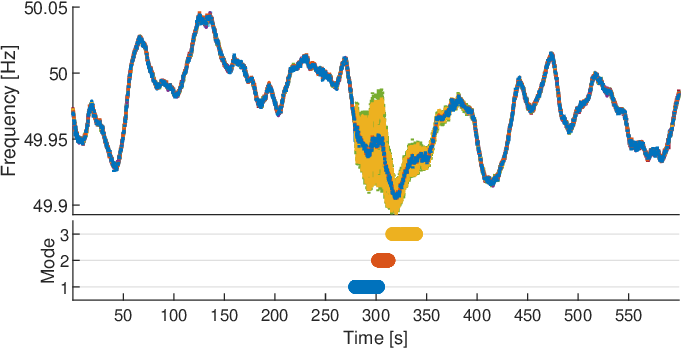}
  \caption{The PMU measurements in the upper plot show oscillations following an event, recorded at 8 different locations. The lower plot shows where in time the observations contributing to the different mode estimates are detected, similarly as for the case with the Kundur Two Area System in Fig. \ref{fig:K2A-TimeSeries}.}
  \label{fig:PMU-TimeSeries}
\end{figure}

\begin{figure}
  \centering \includegraphics[width=0.45\textwidth]{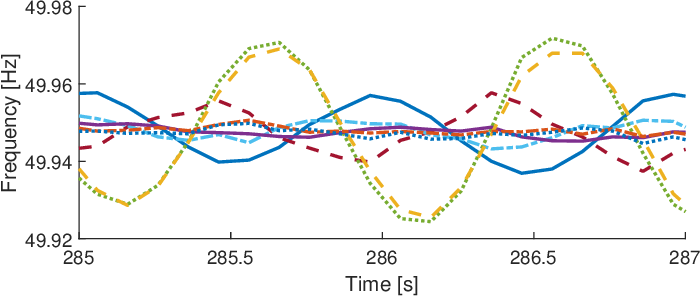}
  \caption{A zoomed view of the original time series are shown. From the period of almost sustained oscillations we see that the method is able to correctly identify the locations with highest oscillations and the phase shift between the various locations. The sampling frequency is \SI{10}{Hz} in this case, which is lower than what can be achieved with most PMUs.}
  \label{fig:PMU-TimeSeries-Excerpt}
\end{figure}

\begin{figure}
  \centering \includegraphics[width=0.45\textwidth]{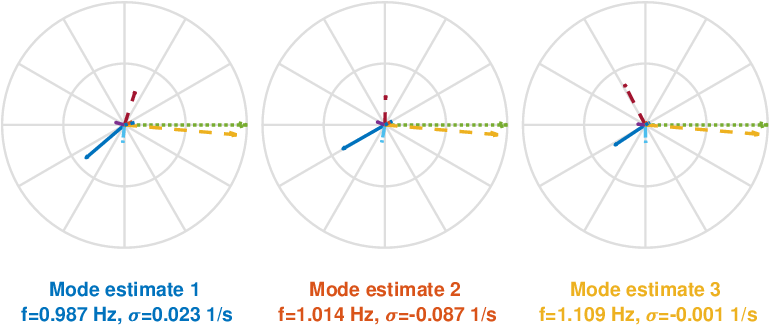}
  \caption{ In the colored captions, the same colors as in Fig. \ref{fig:PMU-TimeSeries} are used to indicate the cluster division, while for the phasors, the colors and line types correspond to those of the markers indicating PMU positions in Fig. \ref{fig:NorwayMap}, and those of the time series in Figs. \ref{fig:PMU-TimeSeries} and \ref{fig:PMU-TimeSeries-Excerpt}.}
  \label{fig:PMU-Modes}
\end{figure}

The recorded PMU data is shown in the upper part of Fig. \ref{fig:PMU-TimeSeries}. Applying the method yields the three mode shapes shown in Fig. \ref{fig:PMU-Modes}. Studying Fig. \ref{fig:PMU-TimeSeries-Excerpt}, which shows a zoomed view at the beginning of the oscillatory period, we see that the oscillations are most prominent in the two signals shown in green and yellow, corresponding to locations S and SW\textsubscript{1} in Fig. \ref{fig:NorwayMap}. These locations are oscillating roughly in phase. Considering locations SW\textsubscript{2} and SE, the amplitudes of the oscillations are a bit lower, and the respective phase shifts are about \SI{135}{\degree} to the left, and \SI{90}{\degree} to the right. Mode Estimate 1 fits well with this  description, indicating that the method is able to produce a reasonable estimate of the mode and corresponding mode shape.

From the lower plot in Fig. \ref{fig:PMU-TimeSeries}, it can be seen that the observations from which Mode estimate 1 is computed, are captured shortly after the oscillations started, Mode estimate 2 was captured shortly after that, and and finally Mode estimate 3 was captured at the end of the period with oscillations. The three mode shapes are similar, but in particular the angle of the phasor corresponding to location SE in Fig. \ref{fig:NorwayMap} (shown with dashed, red lines) changes during the course of the oscillations. Also, the frequencies of Estimates 2 and 3 are a bit higher than the first. This could indicate that some remedial action was taken to mitigate the oscillations.

\section{Discussion}
\label{sec:Discussion}
The above results are promising; first and foremost, the method is able to produce relatively accurate estimates of modes of the power system. In the simulated case, the mode estimates are very similar to analytically calculated modes. In the case with PMU data, the modes are as expected when considering amplitude and phase of oscillations in the time series.

The results from the case with PMU data reveals another potential benefit of the described empirical approach, namely that the method could facilitate tracking of how modes changes during a period. In the presented case, the oscillatory mode and the corresponding mode shape appears to change slightly during the course of the oscillations. This is information that would be difficult to obtain using modal analysis, which could potentially contribute to increased situational awareness.

When applying the method to the Kundur Two Area system, two of the resulting mode estimates are copies of other estimates, appearing at different angular positions. This is caused by the way the observations are stored, i.e. with the longest observability phasor at \SI{0}{\degree}. This could be avoided in a post processing step, checking if any of the mode estimates are replicates of each other, or by storing the observations in a format that is independent of the angle of the mode shape.

Further testing needs to be done to improve the method, and facilitate development into a form that is useful for grid operators. Imaginably, the dynamics decomposition in Part I would run continuously, analyzing numerous PMU data streams on the go, while the clustering in Part II could be performed on a longer-term moving window, for instance of lengths on the scale of minutes or hours. Based on some measure of certainty of each mode estimate (e.g. based on the degree of variance, or the number of observations in the cluster from which it is computed), the estimates could be stored, along with the time of the day when they were obtained, which could further be used to indicate how modes and mode shapes change during the course of load variations throughout a day. This could in turn be taken into account when designing future wide area control applications.

For the implementation, there is a lot of potential for improvement. Other input quantities than the frequency could be tested as input to the method, for instance the voltage angle. Further, in Part I of the method, the criterion for determining whether observations should be included in the clustering part could take many forms. If a clustering algorithm less susceptible to noise was used, the criterion could be less strict, or even be omitted. Other variants of the dynamics decomposition could also be tested, for instance as described in \cite{Messina2011}, where EMD is used to decompose the input signals into IMFs rather than the two-layer combination of PCA and CPCA implemented here.






\section{Conclusion}
\label{sec:Conclusion}
Through testing on simulated data and recorded PMU data, the presented method appears to be a useful tool for digesting numerous PMU measurements into useful knowledge of the system, in the form of information about the oscillatory modes and their mode shapes. The main conclusion is that the method is feasible, and shows promising result on measured and simulated data. Further work is required to improve the different constituents of the method, including the dynamics decomposition using CPCA, the filtering/handling of noise/bad observations and the clustering part. Further work on measured data is necessary to verify the robustness and efficiency of the method for online use.

\bibliographystyle{unsrt} 
\bibliography{references}

\end{document}